\documentclass[prb,twocolumn,showpacs,amssymb]{revtex4}

\usepackage{graphicx}
\usepackage{dcolumn}
\usepackage{bm}

\begin{document}


\title{The effect of trap anharmonicity on the critical
 temperature for Bose-Einstein condensation}
\author{Augusto Gonzalez}
\affiliation{Instituto de Cibernetica, Matematica y Fisica, Calle E
 309, Vedado, Ciudad Habana, Cuba}
\email{agonzale@icmf.inf.cu}

\begin{abstract}
\bigskip
The anharmonicity of a magnetic atomic trap at long distances from 
its center (forty times the radius of the first atomic orbit along
the elongated axis, i.e. hundreds of microns) is shown to increase up to 
45 \% the temperature for Bose-Einstein condensation. This effect is 
perhaps small in the traditional traps, but should certainly be taken into 
account in magnetic microtraps, which characteristic dimensions are of 
order 1 mm.
\end{abstract}

\pacs{03.75.Fi}

\maketitle

Unlike the situation in liquid Helium, atomic vapors undergoing Bose-Einstein
condensation (BEC) in magnetic traps are very rarefied \cite{Legget}. As
a consequence, the free-boson model works extremely well. Interaction
effects, computed in mean-field approximation, are shown to decrease the
critical temperature for BEC in a few percents \cite{Dalfovo}. Still lower
corrections come from the approximation of the one-particle discrete spectrum
by a continuum of states, the so-called ``finite-$N$'' corrections 
\cite{Dalfovo}, which also decrease the critical temperature. 

In the present paper, we show that a relatively important increase of the 
critical temperature (up to 45 \%) could be related to trap anharmonicities at 
``long'' distances from the center. By long distances, we mean around
forty times the radius of the first atomic orbit along the elongated 
direction in the trap potential, i.e. around 200 $\mu$m. In the commonly 
used traps, which characteristic dimensions are a few centimeters, the
belief is that anharmonic effects should be very weak at distances of 200 
$\mu$m. Nevertheless, we present an example of a recent experiment 
\cite{Pereira}, in which the measured values of the total number of atoms 
in the trap and the critical temperature are not consistent unless 
anharmonicity (or other) effects raising $T_c$ are included. On the other 
hand, the trap potential should certainly be anharmonic at 200 $\mu$m in 
the recently developed atomic microtraps \cite{microtraps}, which 
characteristic dimensions are of order 1 mm.

In the continuum free-boson model, the critical temperature is simply
estimated from the conditions $N_0=0$, $\mu=0$ in:

\begin{equation}
N-N_0=\int_0^{\infty} \frac{g(\epsilon){\rm d}\epsilon}
 {\exp(\frac{\epsilon-\mu}{k_B T})-1}.
\label{eq1}
\end{equation}

\noindent
$N_0$ is the number of atoms in the condensate, and $\epsilon\ge 0$ is
the one-atom excitation energy. The chemical potential takes values 
$\mu\le 0$. $g$ is the density of states.

Magnetic traps with cylindrical symmetry are very common. The trap
potential, coming from the interaction of an hyperfine atomic species 
with the magnetic field, is proportional to the latter which, near
the trap center, has a minimum and is written as:

\begin{equation}
B(\rho,z)=B_0+b_{\rho} \rho^2+b_z z^2.
\label{eq2}
\end{equation}

\noindent
The corresponding harmonic oscillator frequencies are $\omega_{\rho}$
and $\omega_z$. It is useful to define $\omega_0=(\omega_{\rho}^2
\omega_z)^{1/3}$, in terms of which the level density is written
$g(\epsilon)=\frac{1}{2} \epsilon^2/(\hbar\omega_0)^3$, and the
critical temperature becomes:

\begin{equation}
k_B T_c^0=\left(\frac{N}{\zeta(3)}\right)^{1/3}\hbar\omega_0=
 0.94~ N^{1/3}~\hbar\omega_0,
\label{eq3}
\end{equation}

\noindent
where $\zeta(x)=\sum_{n=1}^{\infty}1/n^x$ is the Riemann Zeta function.

The corrections to $T_c$ due to interaction or finite-$N$ effects take,
respectively, the form \cite{Dalfovo,Salasnich}:

\begin{equation}
\frac{T_c-T_c^0}{T_c^0}=-1.33~ \frac{a}{r_0}~ N^{1/6},
\label{eq4}
\end{equation}
\begin{equation}
\frac{T_c-T_c^0}{T_c^0}=-0.73~\frac{\bar\omega}{\omega_0} N^{-1/3},
\label{eq5}
\end{equation}

\noindent
where $r_0=\sqrt{\hbar/(m\omega_0)}$, $a$ is the s-wave scattering length,
and $\bar\omega=(\omega_z+2\omega_{\rho})/3$.
Typical values $N\sim 10^6$, $a/r_0\sim10^{-2}$ lead to small numbers in
the r.h.s. of Eqs. (\ref{eq4},\ref{eq5}). As mentioned, the corrections
(\ref{eq4}) are computed within mean-field theory. More elaborated variational
calculations based on diffusion Monte-Carlo \cite{Giorgini} or hypernetted
chain theory \cite{Fantoni} support the validity of mean-field approximations
at the relevant particle densities. Notice that both corrections (\ref{eq4})
and (\ref{eq5}) are small and make $T_c$ decrease. In a general power-law
trap, the corrections may be of any sign and their relative magnitudes may
be still greater \cite{Salasnich}. However, for usual 
magnetic traps one expects a harmonic potential coming from (\ref{eq2}), at
least in the vicinity of the trap center.

\begin{figure}[ht]
\begin{center}
\includegraphics[width=.7\linewidth,angle=-90]{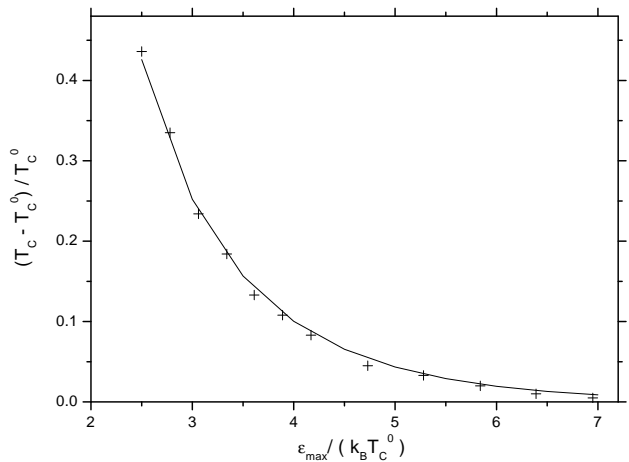}
\caption{\label{fig1}Shift in the critical temperature as a function of
 the energy cutoff. Solid line: corrections coming from Eq. (\ref{eq6}),
 crosses: finite-$N$ calculations for the trap used in Ref. 
 \onlinecite{Pereira}.}
\end{center}
\end{figure}

Let us suppose that at long distances, $r\gg r_0$, anharmonic
terms are relevant in (\ref{eq2}). In fact, one expects a saturation 
of the magnitude of $B$, which should approach the bias field value as the 
distance increases. In a trap potential that saturates at long distances, 
the density of energy levels decreases at high energies\cite{nota}. It 
means that the occupation of the ground state is increased, and thus the 
critical temperature is raised.

We will model the density of levels of the (saturating) anharmonic 
potential in the simplest way: by truncating the one-particle spectrum. It 
means that an upper integration limit, $\epsilon_{max}$, is set in 
(\ref{eq1}). The critical temperature is thus determined from:

\begin{eqnarray}
t^3 \zeta(3)&=&\zeta(3)-Li_3(e^{-tx})-tx~Li_2(e^{-tx})\nonumber\\
 &+&\frac{(tx)^2}{2}\ln(1-e^{-tx}),
\label{eq6}
\end{eqnarray}

\noindent
where we have defined $t=T_c^0/T_c$, $x=\epsilon_{max}/(k_B T_c^0)$,
and $Li_k(z)=\sum_{n=1}^{\infty}z^n/n^k$ is the polylogarithm function.

The corresponding correction to $T_c^0$, i. e. $1/t-1$, is shown in Fig.
\ref{fig1} as a function of $x$. It is around 10\% when $\epsilon_{max}
\approx 4 k_B T_c^0$. As $k_B T_c^0 \sim 10^3 \hbar\omega_z$ in many 
of the experiments, energies of order $4 k_B T_c^0$ involve particle
orbits of radius around 60 $r_{0z}$. On the other hand, when 
$\epsilon_{max}/k_B\approx 2.5~T_c^0$, corrections to $T_c^0$ reach 40 \%, 
and the cutoff distance is around 40 $r_{0z}$. A fit to the long-x tail of 
the curve leads to:

\begin{equation}
\frac{T_c-T_c^0}{T_c^0}=22~ \left(\frac{k_B 
 T_c^0}{\epsilon_{max}}\right)^4.
\label{eq7}
\end{equation}

In quality of example, let us consider the trap used in Ref. 
\onlinecite{Pereira}, in which BEC of spin-polarized Helium was achieved. 
The example is quite interesting because independent measurements of
$N$ and $T_c$ are reported. The authors-provided frequencies are 
$\nu_z=115$ Hz, $\nu_{\rho}=1090$ Hz. The critical temperature, number of 
atoms in the trap and the scattering length are estimated as 
$T_c=4.7\pm 0.5~\mu$K, $N= (5\pm 2.5)\times 10^6$, $a\approx 16$ nm, 
respectively. Turning back to Eqs. (\ref{eq3}-\ref{eq5}), 
we obtain that a temperature $T_c=4.7~\mu$K, corresponds to a 
number of atoms $N=1.4\times 10^7$, well above the error bars. On the 
other hand, if we take the lowest value $T_c=4.2~\mu$K, then the number of 
atoms is $N=8\times 10^6$, a value closer but still outside error bars.
The reason for this apparent discrepancy may be either the rough 
estimation of $N$, or the anharmonicity corrections, which have not been 
included.

In conclusion, we have shown that the saturation of the magnetic field 
at distances around 200 - 300 $\mu$m from the trap center leads to a 
decrease of the density of energy levels, and thus to an increase of the 
ground-state occupation and to an increase of $T_c$. We have modeled the 
decrease of the level density simply by means of an energy cutoff in the 
single-particle spectrum. The corresponding 
correction to $T_c^0$ partially cancels out with interaction or 
finite-$N$ effects, both of which lower the critical temperature. 
Although expected to be small, detailed calculations of the magnetic 
field configurations in the commonly used traps are needed in order to 
evaluate the role of anharmonicity corrections. With regard to the atomic 
microtraps, which characteristic dimensions are of order 1 mm, 
anharmonicities are expected to be very important and should be taken into 
account.

\begin{acknowledgments}
Part of this work was carried out at the Abdus Salam ICTP. The author
acknowledges the ICTP Associate and Federation Schemes for support. 
\end{acknowledgments}


\begin{thebibliography}{99}
\bibitem{nota} We assume that the confinement potential decays abruptly 
 above $r_0$. The density of energy levels 
 refers to states in which the particle motion is confined inside the trap
 volume, i.e. the discrete spectrum. The contribution of states in the 
 continuum spectrum (in which the particle is free to move in all the space)
 to the trap thermodynamic properties is proportional to the probability of 
 finding the particle inside the trap, i.e. equal to zero.
\end{thebibliography}
\end{document}